\documentclass{article}[12pt]
\usepackage{blindtext}
\usepackage[margin=1in]{geometry}
\usepackage{setspace,ulem}

\usepackage{authblk}
\usepackage{graphicx}
\usepackage{xcolor, multirow, natbib, hyperref, amsmath, bm}
\DeclareMathOperator*{\argmin}{arg\,min}

\title{Accounting for Temporal Variability in Functional Magnetic Resonance Imaging Improves Prediction of Intelligence}
\author[1,4]{Yang Li}
\author[2,4,5]{Xin Ma}
\author[1]{Raj Sunderraman}
\author[1]{Shihao Ji}
\author[3]{Suprateek Kundu}
\affil[1]{Department of Computer Science, Georgia State University}
\affil[2]{Department of Statistics, Florida State University}
\affil[3]{Department of Biostatistics, The University of Texas at MD Anderson Cancer Center}
\affil[4]{Co-first authors}
\affil[5]{Corresponding author. Email: xm22a@fsu.edu}
\date{}

\begin{document}

\maketitle

\begin{abstract}
Neuroimaging-based prediction methods for intelligence and cognitive abilities have seen a rapid development in literature. Among different neuroimaging modalities, prediction based on functional connectivity (FC) has shown great promise. The overwhelming majority of literature has focused on prediction using static connectivity, but there are limited investigations on the merits of such analysis compared to prediction based on dynamic FC or using region level functional magnetic resonance imaging (fMRI) times series data that encode temporal variability. To account for the temporal dynamics in fMRI data in prediction pipelines, we propose a deep neural network involving a bi-directional long short-term memory (bi-LSTM) approach that also incorporates an $L_0$ regularization mechanism for feature selection. The proposed deep-learning pipeline involving temporally-varying input features is applied for predicting cognitive scores based on region level fMRI time series data as well as dynamic FC features. We implement the method via an extremely efficient GPU computation framework. We undertake a detailed comparative analysis of prediction performance for different intelligence measures based on static FC, dynamic FC, and region level fMRI time series data acquired from the Adolescent Brain Cognitive Development (ABCD) study involving close to 7000 individuals. Our detailed analysis illustrates that static FC consistently has inferior prediction performance compared to region level fMRI time series data or dynamic FC-based features for unimodal rest and task fMRI experiments, as well as in almost all cases of prediction using a combination of task and rest fMRI features. In addition, the proposed bi-LSTM pipeline based on region level fMRI time series data identifies several shared as well as differential important brain regions across task and rest fMRI experiments that drive intelligence prediction. A test-retest analysis of the selected features shows strong reliability across cross-validation folds.  Given the large sample size from ABCD study, our results provide strong evidence that superior prediction of intelligence can be achieved by accounting for temporal variations in fMRI.
\end{abstract}

\textbf{Keywords:} Intelligence prediction; deep neural networks; neuroimaging analysis; feature selection.

\doublespacing

\newpage
\section{Introduction}

There is great interest in understanding the neural underpinnings of individual differences in intelligence, because it is one of the most important predictors of long-term life success. Intelligence may be measured via cognitive measures that may include fluid intelligence, defined as the ability to use inductive and deductive reasoning (independent of previously acquired knowledge) to solve new problems \citep{kyllonen2017fluid}, or crystallized intelligence that involves knowledge that comes from prior learning and past experiences, among others. There is an ongoing debate on whether variations in such intelligence measures are more correlated with the brain structure, or the brain function, and investigations are underway to discover the optimal set of neuroimaging features that are most predictive of these intelligence levels. However, it is a major challenge to relate structural and functional properties of the brain to complex behavioural expression or function \citep{bullmore2009complex,le2001diffusion,raichle2001default}. Traditional investigations in literature have used structural  neuroimaging-derived features such as whole brain volume, regional gray and white matter volumes or regional cortical volume/thickness and diffusion indices at the whole brain level. While useful, these features may smooth over discriminative features at a finer resolution resulting in inadequate prediction of intelligence \citep{chen2020fluid,paul2016dissociable,ritchie2018much,yuan2018fluid}.  Prediction using structural connectivity data derived from diffusion tensor imaging was proposed by \cite{kawahara2017brainnetcnn} who developed an approach involving convolutional neural networks.

Although traditional literature focused on structural brain measures for predicting intelligence, more recent studies have started to investigate prediction strategies for intelligence based on functional MRI  \citep{shen2017deep,ferguson2017fluid,dubois2018distributed,liu2018chronnectome,kashyap2019individual,he2020deep}. While the majority of intelligence prediction approaches involve linear regression methods, recently emerging studies have focused on non-linear approaches including polynomial kernel SVR \citep{wang2015mri}, and kernel ridge regression \citep{he2020deep,li2019global} methods. A recent review by \cite{vieira2022prediction} involving 37 studies concluded that while there is a plethora of studies showing correlations between brain function and intelligence, there is only a recent emerging trend of predicting intelligence based on functional brain features using machine learning algorithms. A recent paper by \cite{abrol2021deep} systematically showed that deep neural networks when trained on raw data outperform classical linear
and non-linear machine learning models in the prediction of age, gender and Mini Mental State Examination scores. This advantage of the deep learning framework is potentially due to the ability to capture non-linear and complex patterns of relationships between the fMRI time series signals and intelligence scores, which may not be adequately represented by linear models. Extensive numerical studies by \cite{he2020deep} validated that the predictive performance of kernel ridge regression approaches based on FC features was essentially comparable to prediction using deep learning methods. The review by \cite{vieira2022prediction} further concluded that while fMRI was the most often used modality for predicting intelligence, most methods used resting state static functional connectivity (FC) derived from the fMRI time series to predict intelligence \citep{li2019global}. Indeed, FC based differences have been validated in individuals with varying cognitive abilities \citep{hearne2016functional}.  The overwhelming majority of intelligence prediction approaches based on fMRI data has relied on resting state static FC as input features in the prediction model. In addition, there are a handful of recent studies have shown that combining FC from task fMRI and resting state fMRI modalities resulted in superior prediction of intelligence measures \citep{elliott2019general,gao2019combining,chen2022shared} compared to prediction using FC derived from any single fMRI modality. These studies discovered shared and unique network features that were common across tasks and rest, which were able to explain the variations in intelligence.

One major drawback of the prediction pipelines based on static FC is that they do not have the ability to incorporate temporal variations in fMRI which may encode important information regarding brain activity influencing cognition and intelligence. For example, summary measures of blood oxygenation level dependent (BOLD) signal variability were found to be associated with total composite cognitive score \citep{sheng2021coupling}. Static FC features average over such temporal variations that may lead to information loss. There are some limited and recent related literature on prediction approaches for intelligence based on dynamic FC, which capture the temporal variations in the brain network over time. For example, recent work by \cite{sen2020predicting} used tensor decomposition to extract features from dynamic brain networks and subsequently used these features for predicting intelligence in the Human Connectome Project (HCP) study. A handful of other related studies include \cite{liu2018chronnectome} who analyzed data on 105 HCP subjects using summary network features, and \cite{omidvarnia2021temporal} who discovered non-random correlations between temporal complexity of resting state networks and fluid intelligence, based on an analysis of 987 HCP individuals. While useful, these articles used summary measures derived from the dynamic networks that potentially result in information loss, which in turn may compromise prediction performance. Most recently, \cite{fan2020deep} directly used the resting-state dynamic FC features to predict intelligence based on approximately 1200 subjects from the HCP study, using a deep neural network model. 

Apart from the above few articles, there is a scarcity of prediction methods based on dynamic brain networks, which is in sharp contrast to the rich literature identifying potential correlates of task-based dynamic functional connectivity with behavior \citep{thompson2013short,kundu2018estimating,kundu2021developing}. Moreover, in-depth comparisons between the predictive ability of static and dynamic FC are fairly limited in literature, with the exception of the recent work by \cite{sen2020predicting} who provide some insights based on   475 subjects from the HCP study. Their analysis relies on tensor-based feature extraction from the dynamic FC features that were derived from a small number of brain regions (85) extracted using the Freesurfer cortical parcellation atlas \citep{desikan2006automated}. While useful, the chosen atlas may not be best suited for brain network-based comparisons. Indeed, it may be preferable to use more recent atlases such as the one proposed in \cite{gordon2016generation} that uses FC boundary maps to define parcels that represent putative cortical areas with highly homogeneous resting state FC patterns within a given parcel. Moreover, larger sample sizes are desirable for more robust and reproducible results, especially given the high-dimensionality of brain networks. In summary, a detailed large scale analysis comparing predictive abilities of static versus dynamic FC features based on more refined parcellations is warranted and much-needed.

Another important aspect to consider when using dynamic FC features for predicting cognition is that these features are essentially engineered from region level fMRI time series and hence can not capture the temporal variability in brain function that is encoded in the original fMRI data. In fact, the relationship between functional brain connectivity strengths and resting state fMRI temporal complexity changes over time scales \citep{omidvarnia2021temporal}, and hence may not be straightforward.  Therefore, it is not immediately clear whether FC features are primed for optimal intelligence prediction using machine learning pipelines. This is especially true for prediction approaches based on deep neural networks, which can automatically identify suitable representations from minimally pre-processed data (such as fMRI time series) via a succession of hidden layers embedded in an end-to-end learning module. There are other aspects to be taken into consideration when using brain FC features for prediction. Static FC features derived from fMRI data are expected to be sensitive to the choice of the methodology (e.g.: pairwise vs partial correlations) and tuning parameters (e.g.: for controlling sparsity of brain networks), which can affect the ensuing predictive analysis. Similarly, results under dynamic FC analysis can be sensitive to the choice of the window lengths \citep{lindquist2014evaluating}. Moreover, the noise level in the FC features may be unpredictable, which may compromise the quality of the analysis. For example, \cite{tian2021machine} illustrated that the test-retest reliabilty of connectome feature weights is generally poor across a range of predictive models, even when the predictive accuracy itself is moderately high.  Therefore, it is necessary to investigate whether one can directly use the observed region level fMRI time series data for intelligence prediction under a deep learning framework, and if and when that provides any advantages compared to prediction based on engineered brain FC features.

The above arguments lay the groundwork for the unmet need to perform a detailed investigation into the prediction properties based on static FC, compared to region level fMRI time series data  as well as dynamic FC features when modeling intelligence. We note that region level fMRI time series data is naturally more attractive to work with, given that they capture temporal variations in brain activity and are inherently of much smaller dimensions compared to brain networks. Directly using region level fMRI time series data as input features in a deep neural network immediately alleviates several of the issues encountered in the case of network based prediction algorithms. For example, it greatly reduces the number of model parameters, thus ameliorating concerns about over-fitting and bypassing computational challenges that may arise in the presence of tens of thousands of edges in the brain network. It also helps in interpretability as the important brain regions associated with the clinical outcomes can be more easily understood and visualized. In contrast, a network based analysis often reports local (eg: edge-level) network features that can be extremely granular and noisy, or global network features that are often abstract and difficult to interpret. In addition, it is desirable to perform the proposed predictive analysis using deep neural network modeling given their success \citep{abrol2021deep} in neuroimaging studies, and given the fact that it is designed to leverage complex relationships between intelligence scores and fMRI data. For example, a deep neural network using region level fMRI time series as inputs may be able to leverage a succession of hidden layers to capture important associations between brain network features and cognitive outcomes, where the brain network features arise in an automated manner within the deep neural network without having to use brain FC features as inputs.

Motivated by the above discussions, our goals in this article are two-fold. First, we provide detailed analysis to evaluate whether static FC-based predictive approaches have any definitive advantages over prediction using (i) region level fMRI time series data; and (ii) dynamic FC features, derived from resting state as well as task fMRI data. Our in-depth analysis spans multiple types of intelligence measures (fluid, cyrstallized, and total composite intelligence) and leverages multiple fMRI modalities including resting state fMRI and 3 types of task fMRI (MID, SST, nback; see details in Section \ref{subsec:data}) experiments. Most importantly, our analysis is based on a large sample size of slightly less than 7000 individuals from the Adolescent Brain Cognitive Development (ABCD) study. This provides a larger scale analysis compared to the overwhelming majority of existing neuroimaging based prediction studies for intelligence that have primarily focused on data sets with much smaller sample sizes, with the exception of a recent network-based study by \cite{chen2022shared}.  As  a result, the findings in our article are much more robust and expected to be far more generalizable.  In order to produce a successful prediction pipeline involving temporally varying neuroimaging features including region level fMRI time series and dynamic FC, we use a novel deep neural network approach based on bi-directional long short-term memory (bi-LSTM) model \citep{schuster1997bidirectional}. On the other hand, the prediction based on static FC features is implemented via a kernel ridge regression model that has shown good success in multiple neuroimaging studies \citep{he2020deep,chen2022shared}. We implement the bi-LSTM approach via an extremely efficient graphics processing unit (GPU) computation scheme, and make the code public (via Github) for practical usage by the broader community. Our second goal involves identifying shared and unique brain regions in task and rest fMRI whose time-varying activity is significantly associated with intelligence. By focusing on brain function instead of networks, our results provide complimentary findings to recent work aimed at discovering shared and differential brain network features pertaining to task and rest functional connectivity that  drive variations in intelligence \citep{chen2022shared}.  We identify such important brain regions using a feature selection approach involving importance weights that are adaptively learnt via a $L_0$-norm regularization on the input layer of the bi-LSTM pipeline. While there is limited precedent for using $L_0$ penalization for feature selection in imaging genetics studies involving structural MRI data \citep{chen2021sparse}, our proposal is one of the first to adapt this idea to a deep neural network framework involving temporally varying functional neuroimaging features under a bi-LSTM framework.

We note that while the proposed approach has some limited technical resemblance with the recent work by \cite{hebling2021deep}, there are also fundamental differences. First, unlike their method, the proposed approach provides feature selection capabilities via an $L_0$ regularization on the input features, which is of paramount importance in neuroimaging studies and especially critical for our second goal of discovering shared and unique brain regions (pertaining to task and rest fMRI) that are related to intelligence.  Second, in addition to prediction based on region level time series data, we also extend our deep learning framework to predict intelligence using dynamic FC,  which is not considered in  \cite{hebling2021deep}. Third, we investigate the capability of both resting state and task fMRI experiments to predict a slew of intelligence measures, which provides a richer analysis compared to the resting state fMRI based prediction presented in that article. Fourth, our analysis is based on the ABCD cohort involving close to 7000 samples, which is orders of magnitudes larger compared to the much smaller HCP cohort that was used for analysis in \cite{hebling2021deep}. Finally, our scientific focus is distinct, in that, we seek to investigate whether incorporating neuroimaging features with temporal variability (either region level fMRI time series data or dynamic FC features) lead to superior prediction of intelligence compared to the routinely used static FC features, and to discover shared and differential brain regions across fMRI modalities that are related to intelligence. Therefore, our treatise provides a more comprehensive and large scale  investigation into unique perspectives that are not necessarily addressed in current literature, to our knowledge.

\section{Materials and Methods}
\subsection{Data}
\label{subsec:data}
We utilized the preprocessed resting-state and task-based fMRI time series from the ABCD-BIDS Community Collection (ABCC; ABCD Collection 3165: \url{https://github.com/ABCD-STUDY/nda-abcd-collection-3165}). More specifically, the parcellated cifti images under Gordon atlas \citep{gordon2016generation} from ABCC release 1.1.0 were downloaded for all the available individuals, including 9608 for the resting-state session, 8101 for the monetary incentive delay (MID) task session, 7953 for the stop signal task (SST) session, and 7955 for the emotional n-back (nback) task session. We only included the 6835 individuals who had fMRI data for all four experiments in the final analysis to have a fair comparison. The typical numbers of points in the fMRI time series are 822, 890, 740 and 1532 for the MID, SST, nback tasks and resting-state sessions respectively. We also collectively refer to these task and resting-state fMRI experiments as `brain states' in the sequel.

The fMRI data in the ABCD study have been pre-processed with the Human Connectome Project's minimal preprocessing pipeline \citep{glasser2013minimal} and the Developmental Cognition and Neuroimaging (DCAN) Labs resting state fMRI analysis tools \citep{fair2020correction} to get the parcellated cifti images, which include fMRI time series from 333 surface regions as defined by the Gordon atlas \citep{gordon2016generation} and also 19 volumetric regions in the subcortical area, adding up to 352 regions in total. Table \ref{tab:gordon} contains the detailed information of the 13 functional modules assigned to the 333 brain surface regions in the Gordon atlas.

\begin{table}[!htp]
\centering
\resizebox{\textwidth}{!}{
\begin{tabular}{|l|lllll|}
\hline
\textit{Module name} & Auditory & Cingulo Opercular & Cingulo Parietal & Default mode & Dorsal Attention \\
\textit{Abbreviation} & Aud & CO & CP & DM & DA \\
\textit{\# of regions} & 24 & 40 & 5 & 41 & 32 \\
\hline
\textit{Module name} & Fronto Parietal & None & Retrosplenial Temporal & Salience & Sensory Motor hand \\
\textit{Abbreviation} & FP & None & RT & Sal & SMh \\
\textit{\# of regions} & 24 & 47 & 8 & 4 & 38 \\
\hline
\textit{Module name} & Sensory Motor mouth & Ventral Attention & Visual & & \\
\textit{Abbreviation} & SMm & VA & Vis & & \\
\textit{\# of regions} & 8 & 23 & 39 & & \\
\hline
\end{tabular}}
\caption{Information on the 13 functional modules in Gordon atlas}
\label{tab:gordon}
\end{table}

We consider three types of neuroimaging features in our analysis. The first one is region level fMRI time series data as coming directly from the parcellated cifti images described above. The second one is the static functional connectivity matrix data, which is obtained by calculating the pairwise Pearson correlations of the region level fMRI time series. The third and last type of features we consider is the dynamic functional connectivities, obtained by calculating the sliding window pairwise Pearson correlations of the time series, with window size selected at 50 seconds and window stride at 5 as suggested in \cite{sen2020predicting}.

In addition to the imaging data, we also extracted the demographical and neurocognition measurements for the corresponding 6835 individuals from the ABCD release 3.0 \citep{karcher2021abcd} year 1 data. Individual's age is included in all models as a control variable. Our outcomes of interest - the intelligence scores - were derived from the NIH toolbox neurocognition battery for the youth \citep{luciana2018adolescent}. Specifically, we consider three types of intelligence metrics including fluid, crystallized, and total composite intelligences.

\subsection{Schematic structure}
\label{subsec:flow}
\begin{figure}[!htp]
\centering
\includegraphics[width=.9\textwidth]{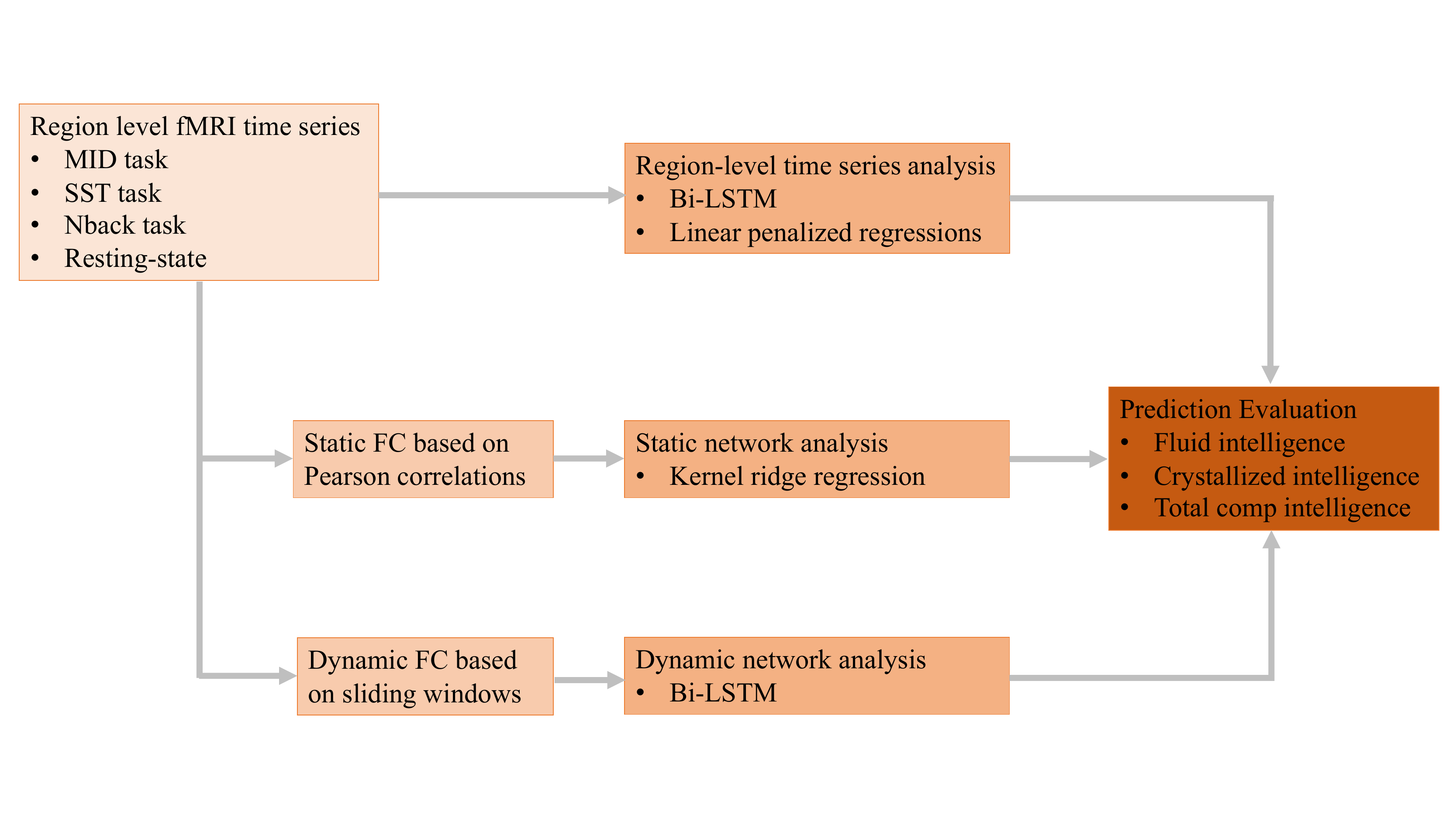}
\caption{Analysis workflow. The static and dynamic FCs are derived from the region level fMRI time series. These neuroimaging features are then served as input for different models for prediction of the three intelligence metrics.}
\label{fig:flow}
\end{figure}

We illustrate the workflow of our analysis in Fig. \ref{fig:flow}. We consider different models for the three types of neuroimaging features extracted from the fMRI images. In the following Sections \ref{sec: bi-lstm} and \ref{sec:feature}, we present our deep neural network prediction pipeline based on temporally varying input features and an integrated feature selection mechanism that can produce feature importance scores. The proposed pipeline can cater to both region level fMRI time-series input data, as well as dynamic FC inputs. Then in Section \ref{subsec:benchmark}, we introduce two types of benchmark models including the kernel ridge regression for prediction based on static FC, and the linear penalized models for prediction based on region level fMRI time series data. In order to evaluate the prediction performance across all the models we consider, we conducted a 5-fold cross validation analysis. We randomly divided the 6835 individuals into 5 partitions. Each time we used a different partition as test data, while the remaining four partitions served as training data. We note that data from different brain states of the same subject was contained either wholly in the training set or the test set. Summary statistics of the five folds are shown in Table \ref{tab:demo}. 

\begin{table}[!htp]
\centering
\resizebox{.95\textwidth}{!}{
\begin{tabular}{  |c  c  c c c c|} \hline
 \multirow{2}{*}{N} & Age &  Sex & Fluid Intelligence & Crystallized Intelligence & Total comp intelligence  \\
 & (months) & (M/F) & (mean $\pm$ sd) & (mean $\pm$ sd) & (mean $\pm$ sd) \\
 \hline
& [107, 132] & 673 / 694 & 92.35 $\pm$ 10.25 & 86.90 $\pm$ 6.81 & 87.01 $\pm$ 8.70 \\
 & [107, 132] & 679 / 688 & 92.72 $\pm$ 10.06 & 87.10 $\pm$ 6.45 & 87.36 $\pm$ 8.41 \\
 1367 $\times$ 5 & [107, 132] & 683 / 684 & 92.39 $\pm$ 10.42 & 86.99 $\pm$ 6.50 & 87.10 $\pm$ 8.54 \\
 & [107, 132] & 693 / 674 & 92.42 $\pm$ 9.99 & 87.16 $\pm$ 6.72 & 87.20 $\pm$ 8.52 \\
 & [107, 132] & 758 / 609 & 92.32 $\pm$ 10.48 & 86.90 $\pm$ 6.61 & 86.99 $\pm$ 8.74 \\
\hline
\end{tabular}}
\caption{Summary on demographic information and intelligence metrics of the five cross-validation folds}
\label{tab:demo}
\end{table}

\subsection{Bidirectional LSTM model}
\label{sec: bi-lstm}
Recurrent Neural Networks (RNNs) have shown great capability in processing sequential data, such as audio, video, traffic data, etc. RNNs contain cyclic connections that repeatedly feed the network activations from a previous step as inputs to the current step, which allows the network to capture long-range dependencies. Long Short-Term Memory (LSTM) \citep{hochreiter1997long} is a specific RNN architecture that introduces control gates: input gates, output gates and forget gates. Compared to conventional RNNs, LSTM alleviates vanishing gradient problems and therefore usually leads to a better performance. Bidirectional LSTM (bi-LSTM) \citep{schuster1997bidirectional} combines an LSTM that takes input in a forward direction with another LSTM that takes input in a backwards direction. Such mechanism allows the model to build dependency among the whole input sequence that is conducive for fMRI data. 

\begin{figure}[ht!]
  \begin{center}
    \includegraphics[width=1\linewidth]{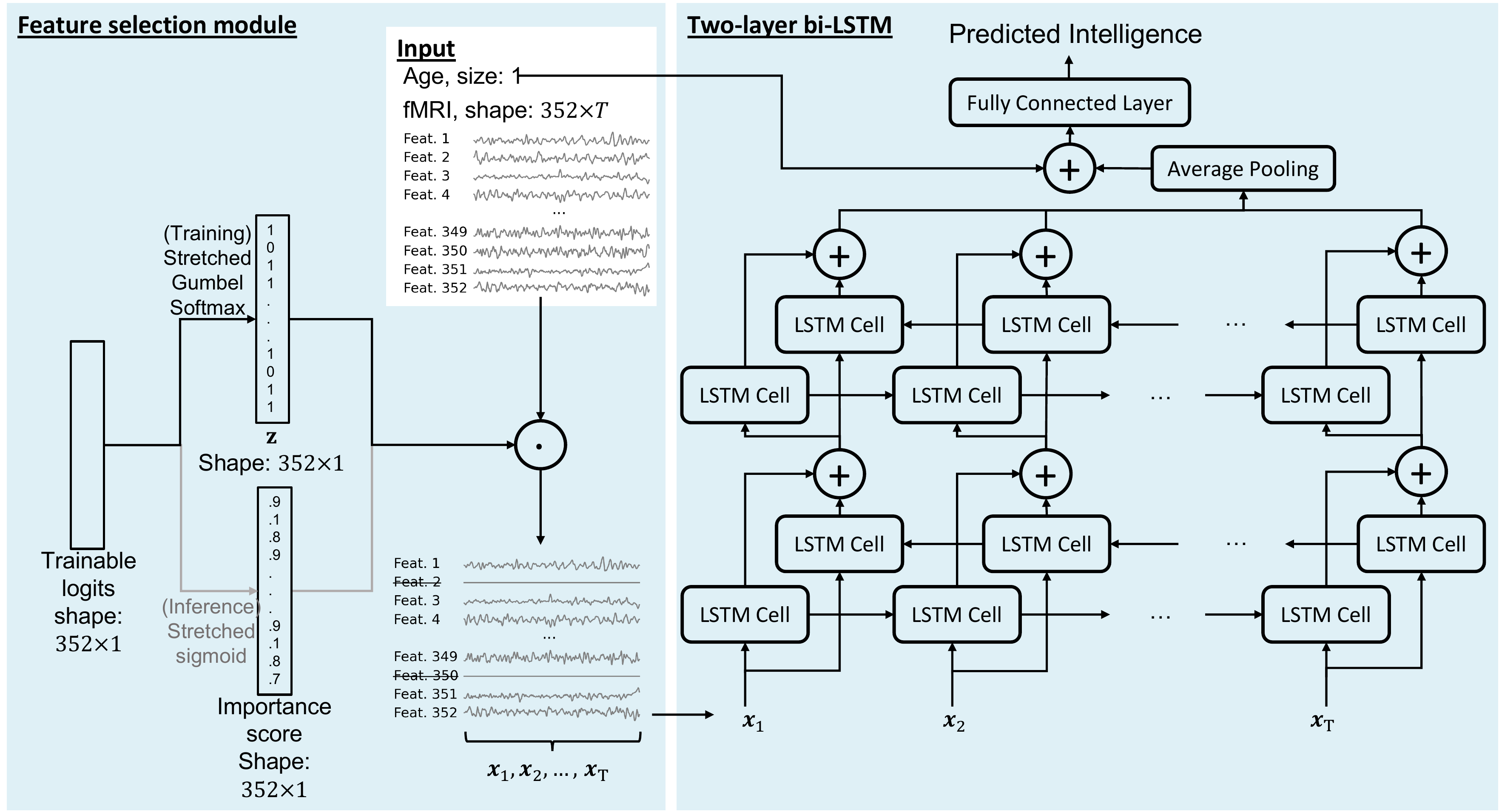}
  \end{center}
  \caption{The architecture of the model trained with feature selection. Left: feature selection module. Right: the architecture of the two-layer bi-LSTM. }\label{fig: arch}
\end{figure}

In our study, we use a two-layer bi-LSTM model. A schematic representation of this model for prediction based on region level fMRI time-series is depicted in Fig. \ref{fig: arch}. As shown on the right side of Fig. \ref{fig: arch}, the model consists of two stacked LSTM modules and a fully connected layer. The input time series data for each individual is a matrix of size $352\times T$, where 352 is the number of cortical surface and sub-cortical regions, and $T$ denotes the number of time points. We denote the column vectors from this data matrix as $\bm{x}_1, \bm{x}_2, ..., \bm{x}_T$, where $\bm{x}_t$ ($t=1,\ldots,T$) is a vector of length 352 for the time point $t$. Inputs $\bm{x}_1$ through $\bm{x}_T$ are fed into the bottom LSTM cells recursively, in both forward and backward directions. The outputs from both directions are concatenated  (denoted as $\oplus$ in Fig. \ref{fig: arch}) and then fed into the top LSTM layer. Between each two layers we use dropout which randomly drops some data in order to prevent overfitting issue. The final prediction is made by a fully connected layer which takes as input the mean aggregated LSTM outputs and the age of the individual, for predicting intelligence.

The bi-LSTM model can be extended to take the dynamic connectivity data as input in predicting intelligence. As these connectivity matrices are symmetric, we only take the upper triangle of each matrix and flatten it as a vector before feeding it into the bi-LSTM model. In this case, the input matrix for the bi-LSTM model is of size $352\times(352-1)/2$ by $T^*$, where $T^*$ is the total number of dynamic matrices, which depends on the window size and window stride when calculating the dynamic connectivities (see Section \ref{subsec:data}). 

We train the model to minimize the difference between the predicted and observed intelligence scores. More specifically, the regularized optimization target is
\begin{equation} 
 \argmin_{\bm{\theta}} \frac{1}{N} \sum_{i=1}^{N} \mathcal{L}(f(\bm{X}_i; \bm{\theta}), y_i) + \lambda \|\bm{\theta}\|_2, \label{eq: LSTM_loss}
\end{equation}
where $f$ denotes the bi-LSTM model parameterized by $\bm \theta$, $\bm{X}_i$ is the matrix of fMRI features (either region level time series, or dynamic connectivities) for individual $i$, $y_i$ is the corresponding observed intelligence score, $N$ is the number of individuals in the training dataset, and $\mathcal{L}(a, b) = (a-b)^2$ is the squared loss function. The second term of Eq. \ref{eq: LSTM_loss} is the $L_2$ regularization (weight decay) on model parameters and $\lambda$ controls the regularization strength. 
 
{\noindent \bf \underline{Implementation:}} We implement the model with PyTorch \citep{paszke2017automatic}. The hidden size of the LSTM is set to 80 for fMRI time series and 500 for dynamic connectivity data. The dropout rate between the two LSTM layers is set to 20\%. We trained the model with backpropagation through time (BPTT) \citep{werbos1990backpropagation} using ADAM \citep{kingma2014adam} as the optimizer. The Adam optimization algorithm is an extension to stochastic gradient descent and has been widely used in deep learning field. For dynamic connectivity data we use ADAM with SAM \citep{foret2020sharpness}. SAM aims to improve model generalization by seeking parameters that lie in neighborhoods having uniformly low loss. We try using SAM in all of our experiments and it turns out that SAM can significantly improve model performance with dynamic connectivity data, which is a novel discovery of independent interest. The learning rate for ADAM optimizer is set to 0.0001 and the weight decay regularization parameter $\lambda$ is set to 0.01. The model is trained for 10 epochs. In terms of validation, we use 5 fold cross-validation as discussed in section \ref{subsec:flow}.

\subsection{Feature selection and importance score learning}
\label{sec:feature}
Feature selection is widely used in machine learning to avoid the curse of dimensionality. By removing or down-weighting the redundant or irrelevant features, issues such as overfitting can be avoided. Additionally, feature selection is naturally equipped to generate feature importance scores and identify the most informative features in prediction tasks, which is of paramount importance in neuroimaging studies. However, feature selection in black box methods such as deep learning is not straightforward, and is an ongoing area of research development. While ad-hoc feature selection methods such as inversion are often used in neuroimaging literature \citep{chen2022shared}, it is desirable to propose more principled and systematic approaches for feature selection  under the deep learning framework.

To enable feature selection in the bi-LSTM model using region level fMRI time series as input, we introduce a binary mask vector $\bm{z}=(z_1,\cdots,z_p)^T$ where $p=352$ is the number of brain regions and $z_k \in \{0, 1\}$ for $k=1,\cdots,p$. When $z_k=0$, the corresponding time series data for region $k$ is zeroed out of the target optimization function. In order to select a small set of important regions, we add into the objection function an $L_0$ regularization term for $\bm{z}$  which explicitly penalizes the number of non-zero $\bm{z}$ and encourages parsimonious models. The optimization target turns into the following format:
\begin{equation} 
 \argmin_{\bm{\theta}, \bm{z}} \frac{1}{N} \sum_{i=1}^{N} \mathcal{L}(f(D_z\otimes\bm{X}_i; \bm{\theta}), y_i) + \lambda_1 \|\bm{\theta}\|_2  + \lambda_2  \|\bm{z}\|_0,  \label{eq: L0_loss}
\end{equation}
where $\otimes$ denotes the matrix multiplication, $D_z$ is a diagonal matrix of size $p$ by $p$ with the elements of $\bm{z}=(z_1,\cdots,z_p)^T$ as its diagonal elements, and $\lambda_1$ and $\lambda_2$ are two hyper-parameters which control the strength of weight decay and feature selection, respectively. However, the binary mask $\bm{z}$ cannot be optimized directly in neural networks since the elements in the vector are discrete variables. Hence, we have to introduce a trainable continuous parameter for $\bm{z}$. A natural choice is to define the elements in $\bm{z} $ as random variables drawn from Bernoulli distribution parameterized by probability vector $\bm{\pi}$. By incorporating the random variables the model becomes stochastic so we have to minimize the expectation over $\bm{z} $ and the optimization target changes into:
\begin{equation} 
\argmin_{\bm{\theta}, \bm{\pi}} \frac{1}{N} \sum_{i=1}^{N} \mathrm{E}_{\text{Bern}(\bm{z} \mid  \bm{\pi})} \left[\mathcal{L}\left(f\left(D_z\otimes\bm{X}_i; \bm{\theta}\right), y_i\right)+\lambda_1 \|\bm \theta\|_2 + \lambda_2 \|\bm{z}\|_0  \right]. \label{eq: stochastic_loss}
\end{equation}
Note that in Eq. \ref{eq: stochastic_loss} we optimize over $\bm{\theta}$ and $\bm {\pi}$. However, the discrete nature of Bernoulli distribution blocks the gradient backpropogation to $\bm \pi$ in the bi-LSTM pipeline. Inspired by sparsity literature in deep learning, we utilize gumbel-softmax \citep{jang2016categorical, maddison2016concrete} as a surrogate for Bernoulli distribution. Gumbel-Softmax was proposed to make categorical variables learnable in neural networks. Instead of sampling the elements of $\bm{z}$ from Bernoulli distribution with probability vector $\bm{\pi}$, we sample a vector $\bm{s}=(s_1,\cdots,s_p)^T$ from the gumbel-softmax distribution as follows:
\begin{equation} 
u \sim \text{Uniform}(0, 1), \quad s_k = \sigma\left( (\log u - \log(1-u) + \alpha_k)/\beta  \right), \quad k=1,\cdots,p,
\label{eqn:gumbel}
\end{equation}
where $\sigma(a)=[e^a/(1+e^a)]$ denotes the logistic function (also known as the Sigmoid function), $\bm{\alpha}=(\alpha_1,\cdots,\alpha_p)^T$ is linked to the probability vector $\bm{\pi}$ through $\pi_k = \sigma(\alpha_k)$ for $k=1,\cdots,p$, and $\beta$ is called the temperature which controls the shape of the distribution. As the temperature approaches 0, $s_k$'s become binary ($0$ or $1$). Note that $s_k$'s cannot be exact 0 or 1 if sampled according to Eq. \ref{eqn:gumbel}, which is not preferable in feature selection because it is not able to completely zero out a feature. Fortunately, \cite{louizos2017learning} proposed a method that stretches the gumbel-softmax samples to the interval of $(\gamma, \zeta)$ where $\gamma < 0 $ and $ \zeta > 1$, and then clamps the samples to be between 0 and 1:
\begin{equation} 
\bar{s}_k=s_k(\zeta-\gamma)+\gamma, \quad z_k^*=\min (1, \max (0, \bar{s}_k)), \quad k=1,\cdots,p
\end{equation}
and then $\bm{z}^*=(z_1^*,\cdots,z_p^*)^T$ can serve as a surrogate of the binary mask vector $\bm{z}$ in the pipeline. \cite{louizos2017learning} provides the conditional expectation of $\|\bm{z}^*\|_0$ given $\bm{\alpha}$ as $\sum_{k=1}^{p} \sigma\left(\bm \alpha_k-\beta \log \frac{-\gamma}{\zeta}\right)$.

Taken all these points together, we substitute in Eq. \ref{eq: stochastic_loss} the Bernoulli distribution with the gumbel-softmax distribution, apply the stretching trick, and turn the optimization target function into this final format:
\begin{equation} 
\argmin_{\bm{\theta}, \bm {\alpha}} \frac{1}{N} \sum_{i=1}^{N} \mathrm{E}_{q(\bm s|\bm{\alpha})} \left[\mathcal{L}\left(f\left(D_g(\bm{s})\otimes\bm{X}_{i}; \bm{\theta}\right), y_i\right)\right] + \lambda_1 \|\bm{\theta}\|_2 +  \lambda_2 \sum_{k=1}^{p} \sigma\left(\alpha_k-\beta \log \frac{-\gamma}{\zeta}\right), \label{eq: final_loss}
\end{equation}
where $q$ is the gumbel-softmax distribution with $\bm{s}=(s_1,\cdots,s_p)^T$ generated as in Eq. \ref{eqn:gumbel} and $D_g(\bm{s})$ is a diagonal matrix with diagonal elements from $g(\bm{s})=\min (1, \max (0, \bm{s}))$. After training the model and obtain the estimates $\hat{\bm{\alpha}}$, we can calculate the feature importance scores as  $\min (\bm{1}, \max (\bm{0}, \sigma(\hat{\bm{\alpha}})(\zeta-\gamma)+\gamma))$.

We show the data flow of the feature selection module on the left panel of Fig. \ref{fig: arch}. The feature selection module is jointly trained with the bi-LSTM module on the right panel. Note that we train the network stochastically by sampling the surrogate mask vector $\bm{z}^*$ from the stretched gumbel-softmax distribution while when generating the importance scores we use the conditional expectation of $\bm{z}^*$ instead. We use the same bi-LSTM model configurations as described in section \ref{sec: bi-lstm}. The model is trained for 20 epochs. We use ADAM as the optimizer with the initial learning rate of 0.0001. The learning rate is multiplied by 0.1 at epoch 10 and 15. In terms of the hyperparameters, we set $\gamma=-0.1, \zeta=1.1$, $\beta=2/3, \lambda_1 = 0.01$, and $\lambda_2=0.25$.

\subsection{Benchmark Comparisons}
\label{subsec:benchmark}
\underline{\noindent Kernel regression methods based on static FC:} We compare the performance of prediction using bi-LSTM approach based on region level fMRI time series and dynamic FC with kernel ridge regression (KRR) based on static FC features that is considered state of the art \citep{he2020deep} - see Appendix A1 of \cite{he2020deep} for more details. The predictors for this approach involve static functional connectivity matrix coming from the resting-state fMRI or any single task fMRI corresponding to unimodal analysis. We also implement the multi-KRR method as described in \cite{chen2022shared} (see supplementary methods S3) for multi-task analysis that involve  static FC features concatenated from the resting-state as well as task fMRI (MID, SST, nback) experiments. As shown in \cite{he2020deep}, the kernel-based methods have comparable prediction performance as a deep learning framework. Thus these single- and multi-KRR methods serve as deep learning benchmark for using static FC as input features. For this reason, we did not train any specific deep learning structure beyond these KRR methods for this type of fMRI features.

\underline{\noindent Linear penalized regression approaches using region level fMRI time series:} We also consider three types of linear regression models using region level time series data, which are fit using penalized approaches to serve as additional benchmark methods for prediction performance. The methods include Lasso with $L_1$-norm penalty \citep{tibshirani1996regression}, ridge regression with $L_2$-norm penalty \citep{hoerl1970ridge}, and elastic net with hybrid $L_1$/$L_2$-norm penalty \citep{zou2005regularization}. The temporal variations in the fMRI time series for each region is summarized into a few leading principal component (PC) scores that account for at least 95\% variation derived from the region level fMRI time series. The principle component analysis is performed separately for each fMRI modality, where 76, 80, 66 and 105 principal components (PCs) are extracted for the MID, SST, nback task and resting-state experiments respectively. The PCs are then stacked into vectors and used as predictors for the linear regression models.

\subsection{Test-retest reliability}
Test-retest reliability is an important indicator of the robustness of the bi-LSTM model with importance score reporting \citep{tian2021machine}. Essentially, a strong test-retest reliability for the features included in the model indicates the robustness of the approach, in terms of being able to consistently identify the important neuroimaging features driving the prediction of the outcome variable. In order to investigate the rest-retest reliability under the proposed approach using region level fMRI time series data, we compute the intraclass correlation coefficient (ICC) metric across all features and cross-validation folds, which captures the agreement across folds in terms of importance rating of the features. We choose to report the ICC of absolute agreement using two-way random effects model with ``average rater" unit \citep{koo2016guideline}. As the raw importance scores are obtained with different selection of regularization parameter under different experiments, they may not be on the same scale between 0 and 1. Thus before computing the ICC scores, we perform a normalization process on each set of importance scores $(z_1,\cdots,z_p)$ with the following formula for $k=1,\cdots,p$:
\begin{equation*}
\tilde{z}_k = \frac{z_k - z_0}{z_1 - z_0},\quad z_1 = \max\{z_1,\cdots,z_p\},\quad z_0 = \min\{z_1,\cdots,z_p\}  
\end{equation*}

\section{Results}
\label{sec:results}

\subsection{Prediction performance}
The evaluation criteria for prediction performance are based on normalized MSE (NMSE) and Pearson correlation (corr) between predicted and observed intelligence scores in the testing samples, which can be calculated as 
\begin{eqnarray*}
 \textrm{NMSE} =  \frac{\sum_{i=1}^{N}(y_i-\hat{y_i})^{2}}{\sum_{i=1}^{N}(y_i-\bar{y})^{2}}, \mbox{  and } \quad \textrm{corr} =\frac{\sum_{i=1}^{N}\left(\hat{y}_{i}-\bar{\hat{y}}\right)\left(y_{i}-\bar{y}\right)}{\sqrt{\sum_{i=1}^{N}\left(\hat{y}_{i}-\bar{\hat{y}}\right)^{2} \sum_{i=1}^{N}\left(y_{i}-\bar{y}\right)^{2}}},
\end{eqnarray*}
respectively,  where $N$ is the total number of individuals in the testing samples, $y_i$ and $\hat{y_i}$ denote the observed and predicted intelligence scores, and $\bar{y}$ and  $\bar{\hat{y}}$ denote the mean of observed and predicted intelligence scores.

The results presented in Tables \ref{tab:lstm_krr}-\ref{tab:linear} capture the comprehensive analysis performed on the ABCD data. The evaluation metrics are averaged over the five cross-validation folds. We compare the prediction performance of the deep learning pipeline using region level fMRI time series, with that using dynamic FC, along with the KRR approaches that use static FC.  Our analysis involves MID, SST and nBack tasks along with resting state fMRI data. Our goal is to investigate which prediction pipeline performs the best, and how the relative performance changes across intelligence metrics and varies by the fMRI modality used for prediction. Unfortunately due to a heavy computational burden, we were unable to provide the results for prediction using dynamic FC features derived from concatenating the three tasks and resting state fMRI experiments.

\begin{table}[!htp]
\centering
\begin{tabular}{| c | c c | c c | c c|} \hline
Type of  & \multicolumn{2}{c|}{Fluid intelligence} &  \multicolumn{2}{c|}{Crystallized intelligence} & \multicolumn{2}{c|}{Total comp intelligence} \\
fMRI experiment & NMSE & $corr$ & NMSE & $corr$ & NMSE & $corr$ \\\hline
\multicolumn{7}{|c|}{bi-LSTM model with region level fMRI time series} \\
\hline
MID & 0.806* & 0.448* & 0.747 & 0.526 & 0.693* & 0.559* \\
SST & \bf{0.844} & \bf{0.403} & 0.798* & 0.468* & \bf{0.756} & \bf{0.500}  \\
nback & 0.786* & 0.472* & 0.760 & 0.511 & 0.676* & 0.575*  \\
rest & 0.835* & 0.407* & 0.778 & 0.484 & 0.733* & 0.517*  \\
3 tasks + rest & 0.775 & 0.483 & 0.704 & 0.554 & 0.652 & 0.594  \\ \hline
\multicolumn{7}{|c|}{bi-LSTM model with dynamic FC} \\
\hline
MID & 0.816 & 0.447 & \bf{0.690} & \bf{0.568} & 0.689* & 0.563*  \\
SST & 0.870* & 0.389* & \bf{0.765} & \bf{0.489} & 0.770* & 0.492*  \\
nback & 0.802 & 0.469 & \bf{0.714} & \bf{0.542} & 0.673* & 0.576*  \\
rest & 0.848 &	0.409 & {\bf 0.719} & {\bf 0.536} & 0.728* &	0.528*  \\ \hline
\multicolumn{7}{|c|}{KRR model with static FC} \\
\hline
MID & 0.840 & 0.401 & 0.752 &	0.498 & 0.754 &	0.496 \\
SST & 0.892 &	0.328 & 0.819 &	0.426 & 0.825 &	0.418 \\
nback &0.810 & 0.439 &0.762 &	0.492 &0.732 &	0.523 \\
rest &0.870 &	0.363 &0.775&	0.478 &0.793&	0.458  \\
3 tasks + rest & 0.768 & 0.482 & \bf{0.660} & \bf{0.583} & 0.650 & 0.592  \\\hline
\end{tabular}
\caption{Prediction performance of bi-LSTM and KRR models across fMRI experiments and intelligence metrics. bold: significantly better performance than the other two models. *: significantly better performance than the KRR model only.}
\label{tab:lstm_krr}
\end{table}

\begin{table}[!htp]
\centering
\begin{tabular}{ |c | c c | c c | c c|} \hline
 & \multicolumn{2}{c|}{Fluid intelligence} &  \multicolumn{2}{c|}{Crystallized intelligence} & \multicolumn{2}{c|}{Total comp intelligence} \\
& NMSE & $corr$ & NMSE & $corr$ & NMSE & $corr$ \\\hline
\multicolumn{7}{|c|}{Lasso} \\
\hline
MID & 0.997 & 0.059 & 0.998 & 0.050 & 0.995 & 0.078  \\
SST & 0.996 & 0.068 & 0.995 & 0.073 & 0.993 & 0.094  \\
nback & 1.000 & 0.000 & 1.001 & 0.000 & 1.001 & 0.000  \\
rest & 0.993 & 0.090 & 0.997 & 0.053 & 0.990 & 0.108 \\
3 tasks + rest & 0.993 & 0.095 & 0.996 & 0.077 & 0.989 & 0.119 \\
\hline
\multicolumn{7}{|c|}{Ridge} \\
\hline
MID & 0.998 & 0.044 & 0.999 & 0.036 & 0.998 & 0.046  \\
SST & 0.997 & 0.052 & 0.999 & 0.030 & 0.997 & 0.056  \\
nback & 0.999 & 0.026 & 1.001 & 0.009 & 1.000 & 0.024  \\
rest & 0.996 & 0.058 & 0.998 & 0.038 & 0.996 & 0.063 \\
3 tasks + rest & 0.995 & 0.070 & 0.998 & 0.043 & 0.995 & 0.072 \\
\hline
\multicolumn{7}{|c|}{Elastic net} \\
\hline
MID & 0.997 & 0.053 & 0.998 & 0.046 & 0.995 & 0.082  \\
SST & 0.996 & 0.059 & 0.995 & 0.069 & 0.993 & 0.090  \\
nback & 0.999 & 0.020 & 1.000 & 0.000 & 1.001 & 0.020  \\
rest & 0.995 & 0.075 & 0.998 & 0.046 & 0.990 & 0.100 \\
3 tasks + rest & 0.994 & 0.083 & 0.995 & 0.078 & 0.989 & 0.114 \\
\hline
\end{tabular}
\caption{Prediction performance of linear penalized models based on region level fMRI time series data}
\label{tab:linear}
\end{table}

When predicting fluid intelligence, the bi-LSTM method with region level fMRI time series has the best predictive performance consistently across all the fMRI task modalities as well as for resting state fMRI (see Table \ref{tab:lstm_krr}). This prediction performance  is significantly better compared to the KRR approaches using static FC. When predicting crystallized intelligence, the bi-LSTM pipeline using dynamic FC have the best prediction performance, which is considerably better than prediction using the bi-LSTM pipeline using region level fMRI data as well as the KRR methods using static FC. Such a superior performance is seen across the three fMRI task modalities as well as resting state fMRI. For the total composite intelligence, the prediction performance based on the bi-LSTM methods using region level fMRI time series data is superior to that based dynamic FC corresponding to the SST fMRI modality, but are similar and comparable for the other fMRI modalities. In contrast, the KRR methods using static FC from task and resting state fMRI data has poor performance when predicting the total composite intelligence score, that is significantly inferior to the bi-LSTM prediction pipelines based on temporally varying input features. Taken together, the above analysis clearly illustrates that using temporally varying fMRI features in a bi-LSTM prediction pipeline has superior predictive performance compared to prediction using static FC with any single task modality and resting state fMRI data.

Further, the prediction accuracy under the MID and the nback tasks was consistently higher compared to resting state fMRI time series data as well as the SST fMRI experiment across all three intelligence metrics, which is consistent with the findings pertaining to static FC based prediction in \cite{chen2022shared}. Moreover, the prediction performance based on combining the 3 tasks and rest fMRI data is comparable under the bi-LSTM pipeline using the region level fMRI time-series input features and the multi-KRR model based on static FC features, when predicting fluid intelligence and total composite intelligence. In general, we note that combining data from multiple brain states may lead to improvements in prediction compared to analysis based on just one fMRI experiment, which is also consistent with the observations in \cite{chen2022shared}.

For crystallized intelligence, the prediction performance under the KRR approach combining all brain states is superior to the bi-LSTM framework based on region level fMRI time-series data. However given the superior prediction performance for crystallized intelligence under the bi-LSTM framework using dynamic FC features based on a single brain state, we conjecture that the prediction under this approach will also be superior compared to the KRR approach based on static FC, when combining data from all brain states. Unfortunately, we were unable to validate this conjecture due to the high computational burden that would be accrued when combining dynamic FC features from all brains states under the bi-LSTM pipeline. Nonetheless, the above analysis provides a thorough comparison and a much deeper insights into the merits of prediction using temporally varying fMRI features compared to prediction based on static FC features.

Finally, we also evaluate the usefulness of the deep learning prediction approach compared to a more routinely used linear model-based prediction using penalization (see Table \ref{tab:linear}) based on fMRI time series features. The results clearly indicate a dismal performance under the linear regression approaches consistently across all task modalities as well as resting-state fMRI data. These results suggest that intelligence prediction using flexible non-linear models provide a far superior performance compared to routinely used linear regression modeling based on region level fMRI time series.

\subsection{Comparison on computation efficiency}
\label{subsec:comptime}
\begin{figure}[!htp]
\centering
\includegraphics[width=.6\textwidth]{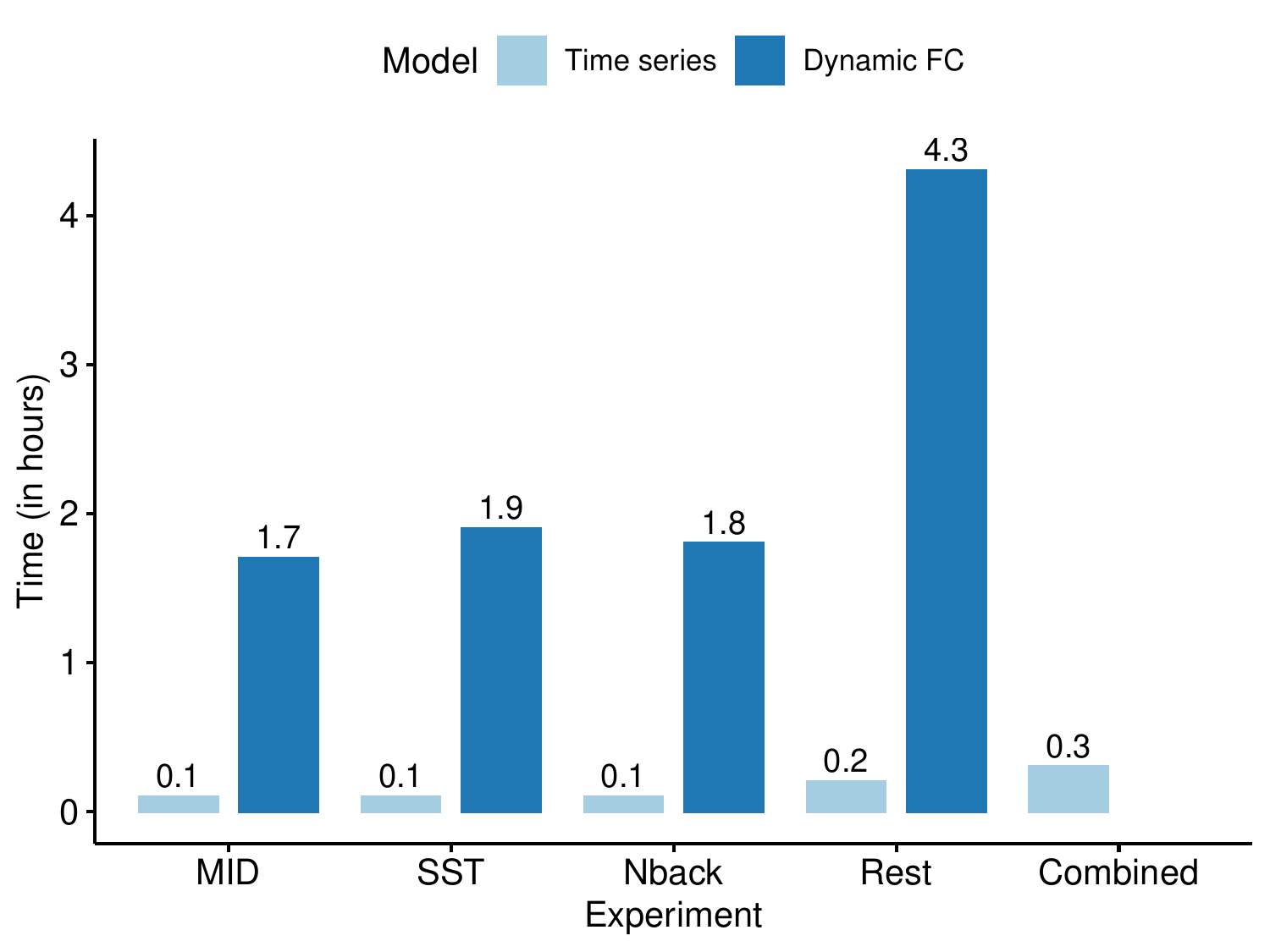}
\caption{Computing time report for bi-LSTM models.}
\label{fig:comptime}
\end{figure}

We report the computation times for bi-LSTM models in Fig. \ref{fig:comptime}. The bi-LSTM models are executed on GPU machine with Intel Xeon Gold 6242 CPU at 2.80GHz and NVIDIA Tesla V100 GPU. We can see from Fig. \ref{fig:comptime} that the training speed for bi-LSTM model with fMRI time series data is typically around 20 times faster compared to the speed with dynamic FC data. The dynamic FC data has much higher dimensional features as input that encumbers the training efficiency and increases the computing time, and at the same time requires greater memory. In contrast, the rapid computation times for prediction using fMRI time series data contribute to its appeal for predicting intelligence.

The KRR and linear penalized models are executed on high performance computing (HPC) environment with Intel Xeon CPU at 2.80 GHz. For KRR model with single task or resting state data, the training time is around one day, while for multi-KRR model with 3 tasks plus resting state data, the training time can last several days. Compared to these computation runtimes in the HPC setting, the bi-LSTM analysis pipeline utilizing GPU computing shows a clear advantage in terms of computational efficiency and scalability. Further, the linear penalized models can be trained quickly, in under five minutes for Lasso and ridge penalty models, and under one hour for elastic net penalty model. The quick computation times result from efficient implementation of these linear models via existing software packages, as well as due to the use of principle components summarizing the variability in the fMRI time series, instead of directly using the original time series data. However, the lack of prediction power rules out this type of model as a viable model for predicting intelligence using region level fMRI time series data.

\subsection{Test-retest reliability of the importance scores}
\label{subsec:icc}
\begin{figure}[!htp]
\centering
\includegraphics[width=.6\textwidth]{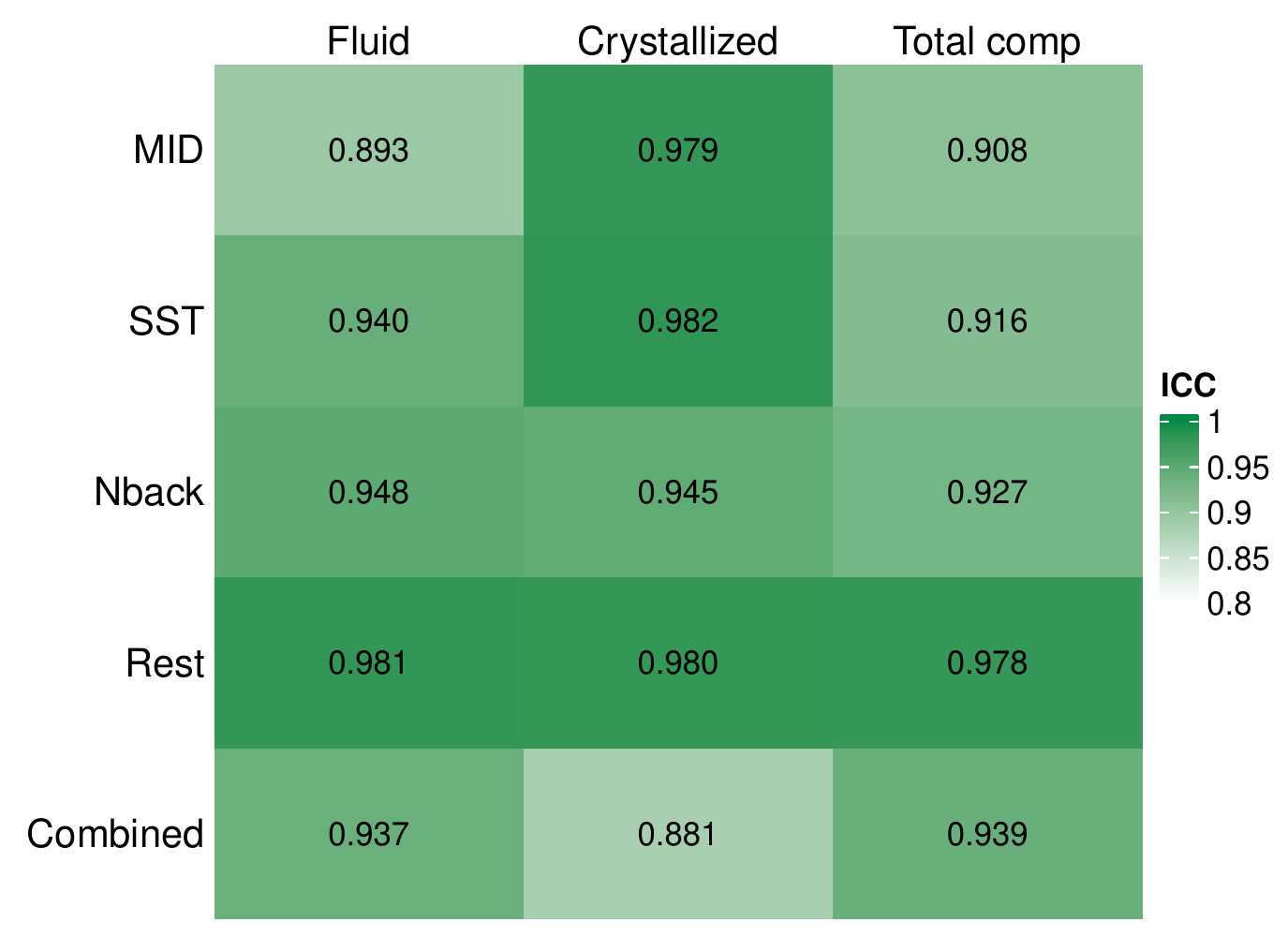}
\caption{Test-retest reliability in ICC scores}
\label{fig:icc}
\end{figure}

We report the summary of ICC scores calculated from the cross-validation folds in Fig. \ref{fig:icc}, based on the scaled importance scores obtained by fitting the bi-LSTM models with $L_0$ regularization using region level fMRI time series data in Section \ref{sec:feature}. From the results, it is clearly evident that the proposed approach has strong test-retest reliability as indicated by high ICC scores, consistently across the three task modalities and resting state fMRI. The important brain regions identified by the proposed approach under the region level fMRI time series analysis is hence highly consistent across the five folds considered in the analysis, and this is true for predicting fluid, crystallized, and total composite intelligences. We note that it was not possible to report test-retest reliability under the other two FC based prediction approaches. This is due to the fact that the KRR method is not naturally equipped to perform feature selection without resorting to additional inversion techniques \citep{chen2022shared}, while the bi-LSTM approach involving dynamic connectivity features has an ultra-high dimensional feature space, making it computationally challenging to implement the $L_0$ regularization mechanism. In contrast, the region level fMRI analysis naturally lends itself to the computation of importance scores via the $L_0$ regularization approach in Section \ref{sec:feature}, in a scalable manner that facilitates the selection of important brain regions that are more interpretable compared to brain network based features. 

\subsection{Brain region selection based on the importance scores}
We have shown in the above Section \ref{subsec:icc} that the scaled importance scores are highly consistent across the five cross-validation folds based on the ICC metric with ``average rater'' unit. This strongly validates our practice to average the importance scores across the five folds and rank the importance of the different brain regions according to the averaged scores. 

\begin{figure}[!htp]
\hspace{-.5in}\includegraphics[width=1.1\textwidth]{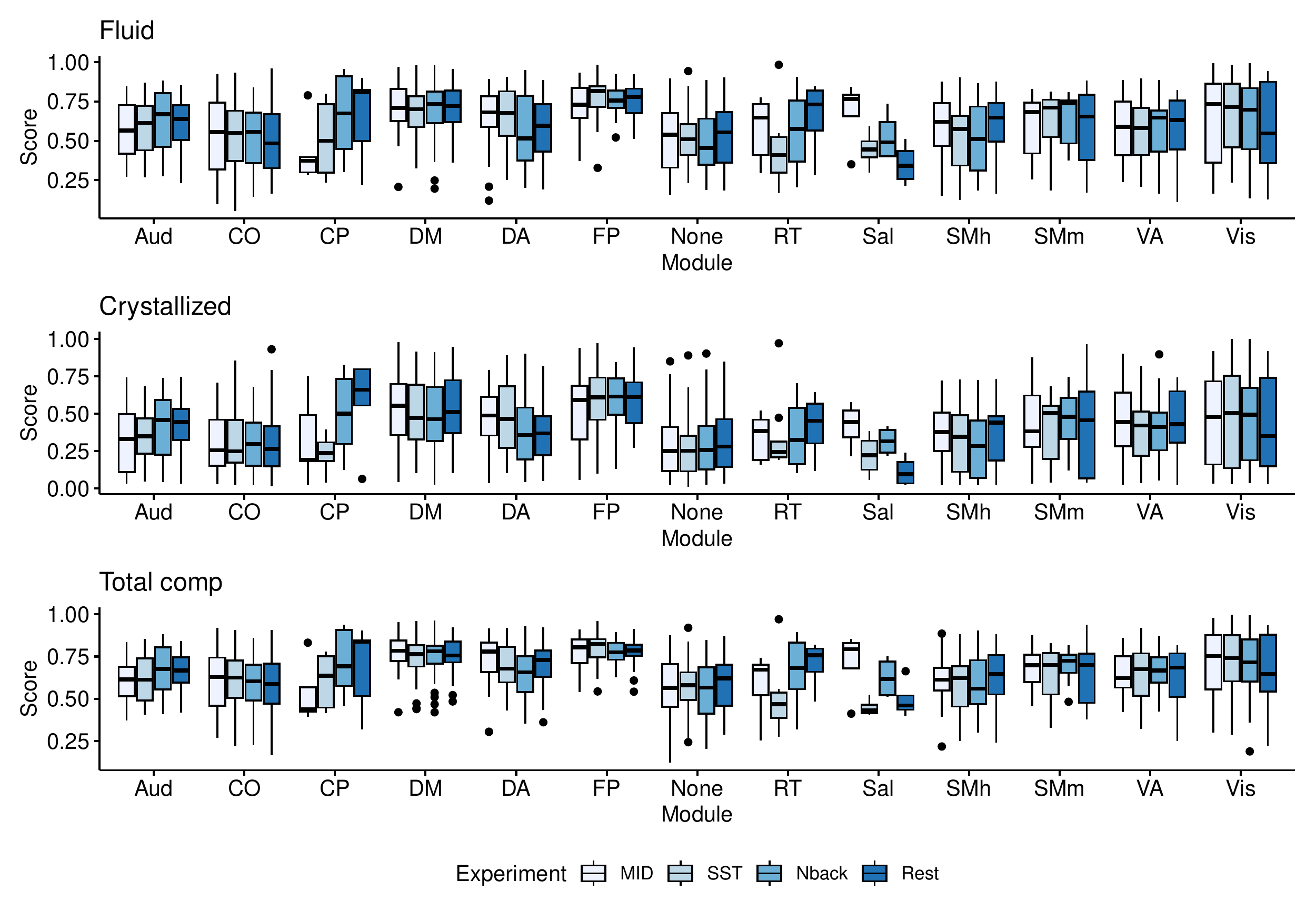}
\caption{Importance scores by functional modules. Refer to Table \ref{tab:gordon} for complete module names.}
\label{fig:imp_module}
\end{figure}

{\noindent \bf \underline{Distribution of Important Brain Regions:}} The overall distribution of importance scores across functional modules are illustrated in Fig. \ref{fig:imp_module} containing boxplots. The fronto-parietal (FP) module and default mode network (DM) have some of the highest importance scores across the three intelligence metrics and brain states, although some FP and DM regions may have low importance for crystallized intelligence. Moreover, cingulo parietal (CP) regions show high importance corresponding to rest and nback tasks, but low importance corresponding to the MID and SST tasks, when predicting intelligence. The regions in the dorsal attention (DA) module have high importance corresponding to the total composite intelligence, but these regions show relatively lower importance when predicting fluid and crystallized intelligence. Regions in the retrosplenial temporal (RT) module show low importance corresponding to the SST task as compared to other brain states, when predicting all the three intelligence metrics. Moreover, regions in the salience (Sal) network have considerably higher importance scores corresponding to the MID task, but not for the other brain states. Finally, the visual network registered high importance scores across brains states and intelligence metrics, although the relative importance of brain regions in this module was higher when predicting fluid and total composite intelligence but slightly lower when predicting crystallized intelligence.

\begin{figure}[!htp]
\centering
\includegraphics[width=.65\textwidth]{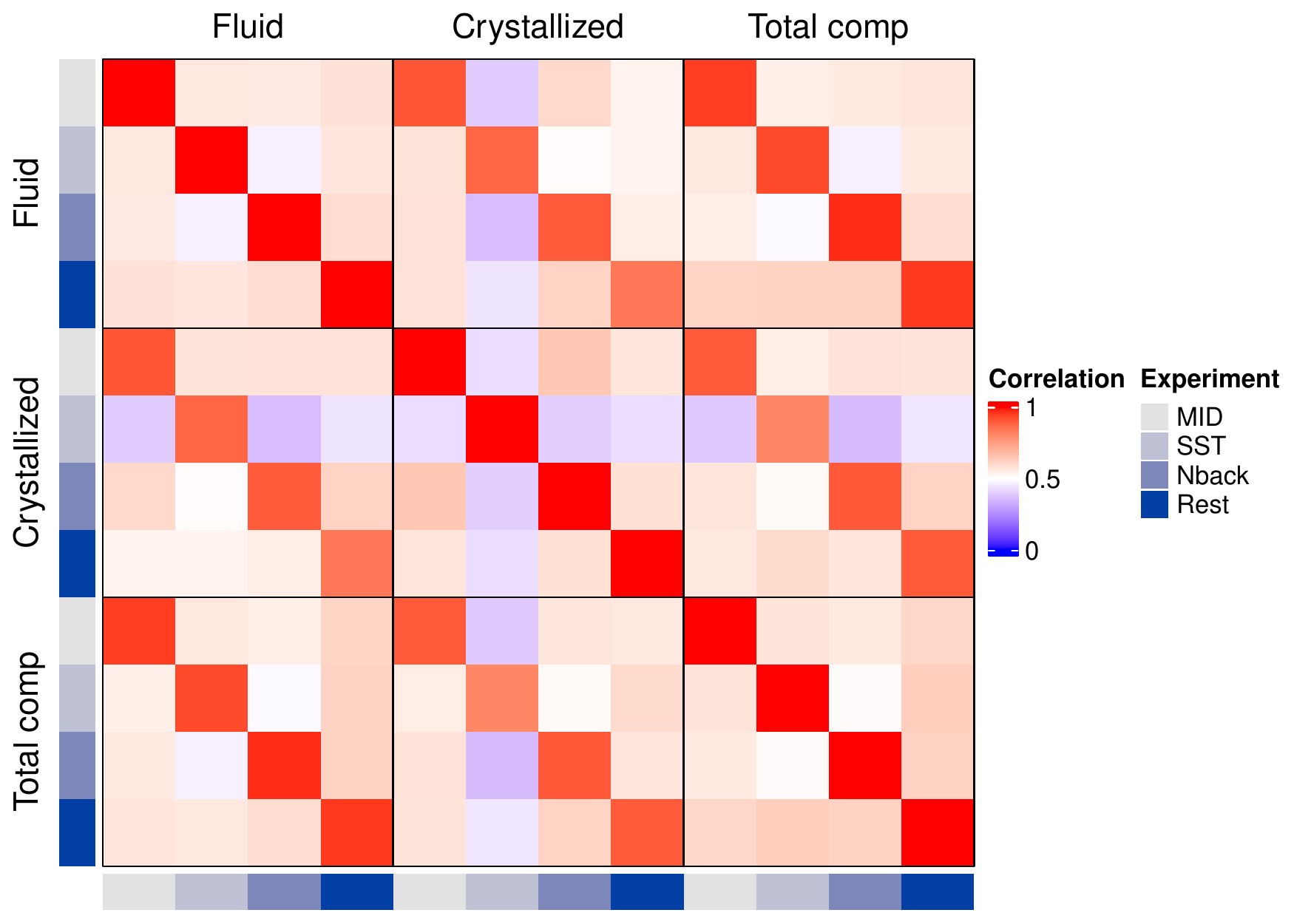}
\caption{Correlation matrices of importance scores across intelligence metrics and brain states. Diagonal panels represent correlation of importance scores across brain state for predicting a given intelligence metric. Off-diagonal panels represent correlation between brain states when predicting different intelligence metrics.}
\label{fig:imp_heat}
\end{figure}

{\noindent \bf \underline{Shared and Differential Brain Regions across Brain States:}}
Fig. \ref{fig:imp_heat} illustrates the correlations of the averaged importance scores across brain states, and intelligence metrics. The figure contains 9 panels, that illustrate correlations between the importance scores across different brain states when (i)  predicting one particular intelligence metric (diagonal panels); and (ii) predicting two distinct intelligence metrics (off-diagonal panels). As evident from the diagonal panels, the correlation between the importance scores across brain states is the highest when predicting fluid intelligence, followed by total composite score and crystallized intelligence, in that order. Further, the feature importance correlations between the SST and nback tasks are seen to be lower when predicting all the three intelligence metrics, with the correlations being the lowest for crystallized intelligence prediction. In general, the  important brain regions from the SST task that drive intelligence prediction have lower concordance with the regions corresponding to other tasks and resting state. In contrast, when predicting total composite intelligence, the importance scores corresponding to brain regions for resting state has strong correlations with other brain states. These results suggest the presence of shared as well as distinct brain regions across brain states that drive intelligence prediction. 

In addition, a close examination reveals that for a given brain state, the importance scores for predicting different types of intelligence metrics show high correlations (off-diagonal panels). This implies the potential for transfer learning under the proposed approach where the fitted model for predicting a particular intelligence metric can be used for predicting another type of intelligence score based on the same type of fMRI experiment (brain state), which is of increasing interest in neuroimaging studies \citep{schirmer2021neuropsychiatric}.

To further investigate shared and differential brain regions in greater detail, we calculated the number of brain regions that were uniquely identified as important for a given brain state but not others, as well as the number of brain regions that were identified as important jointly across pairs of brain states. Moreover, we considered a brain region to be important if it belong to the top 10\% of the importance scores. For greater interpretability, our analysis was stratified by functional modules, and by different intelligence metrics.

\begin{figure}[!htbp]
\hspace{-.5in}\includegraphics[width=1.2\textwidth]{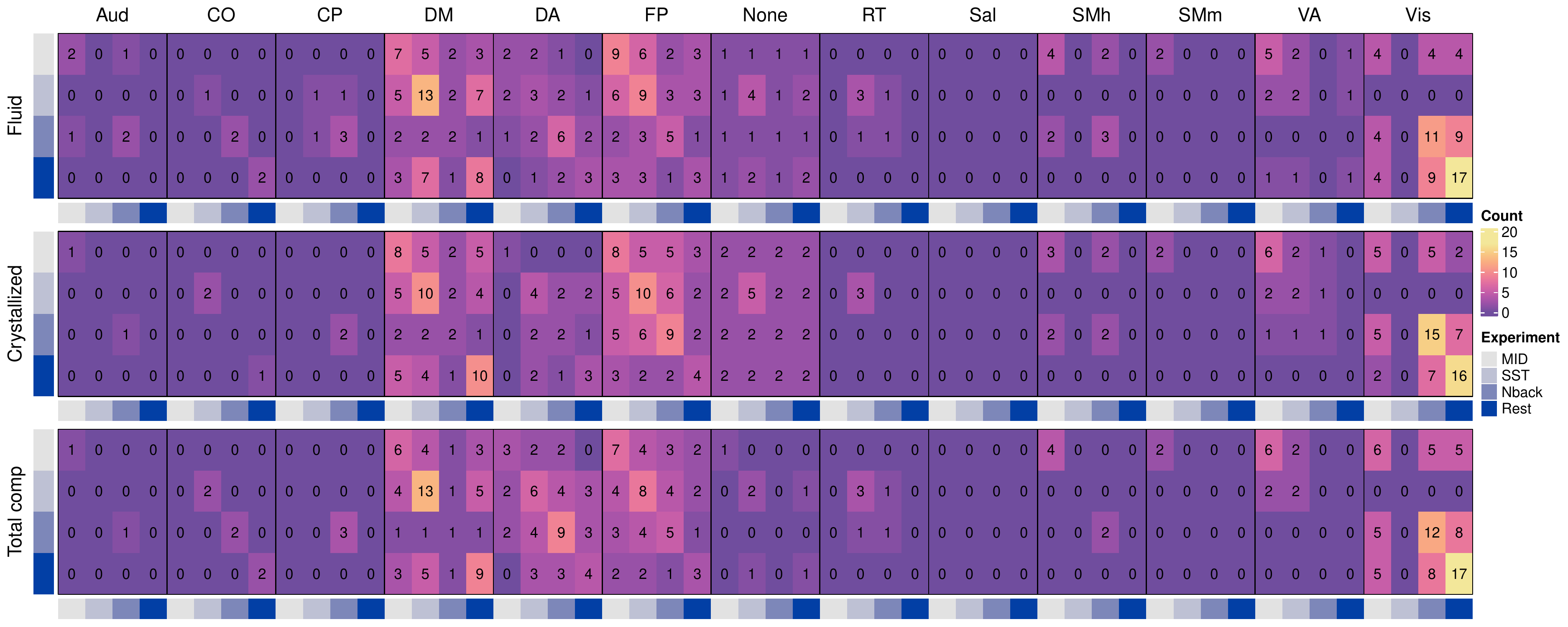}
\caption{Shared and differential top 10\% brain region counts by functional modules. Diagonal panels for each functional module indicate number of unique brain regions under a given fMRI experiment that are important for predicting intelligence. Off-diagonal panels indicate number of shared brain regions that are common across pairs of fMRI experiments when predicting intelligence. }
\label{fig:top_module}
\end{figure}

These results were illustrated in Fig. \ref{fig:top_module}, which presents the number of unique regions in the diagonal cells and shared regions across pairs of brain states in the off-diagonal cells, when predicting intelligence. The highest number of unique brain regions were discovered in (i) the default mode network for all brains states except the nback task; (ii) the fronto-parietal network corresponding to all the brain states except the resting state; and (iii) the visual network corresponding to nback task and resting state. Further, the ventral attention network contained a number of brain regions that were uniquely important corresponding to the MID task when predicting intelligence. In addition, shared brain regions were also discovered. In particular: (i) resting state and nback task shared a high number of important regions in the visual network; (ii) MID and SST tasks and resting state shared common important brain regions in the default mode network; and (iii) several shared brain regions were discovered in the fronto-parietal region that were common pairwise for the the MID, SST, and nback tasks.

\begin{figure}[!htp]
\centering
\includegraphics[width=.8\textwidth]{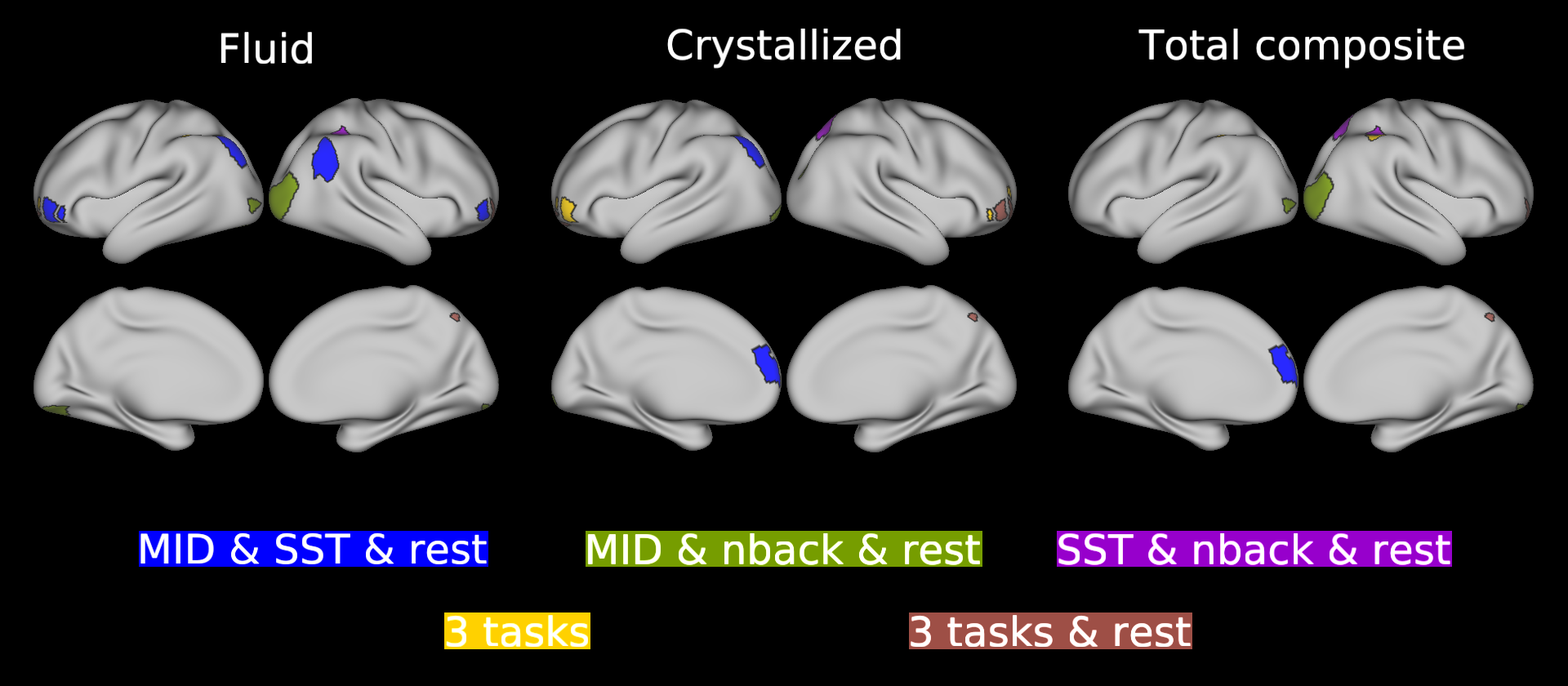}
\caption{Brain maps for common top 10\% brain regions that were shared across 3 or more brain states}
\label{fig:brain_common}
\end{figure}

The shared brain regions common across three of more brain states are presented in Table \ref{tab:brain_common}, and visualized in Fig. \ref{fig:brain_common}. The common brain regions shared across the three tasks lie in: (i) the DMN (2 ROIs), the FP (1 ROI) and the DA network (1 ROI), when predicting fluid intelligence; (ii) in the FP (4 ROIs), the DMN (2 ROIs), the VA network (1 ROI), when predicting crystallized intelligence; and (iii) in the DA network (1 ROI), the FP (2 ROIs) and the DMN (1 ROI), when predicting the total composite intelligence. When comparing shared regions between two tasks and rest for intelligence prediction, these regions were distributed in: (i) the visual and VA network, along with regions in the DMN when predicting fluid intelligence; and (ii) in the DMN and the visual network when predicting crystallized intelligence. The shared regions between two tasks and rest when predicting total composite intelligence were found to be overlapping with the discoveries for fluid and crystallized intelligence prediction, except for one region: region 167, which was assigned to the fronto-parietal module and lies in the right supramarginal gyrus. Finally, regions 257 belonging to the DMN, and 277 belonging to the FP network, along with region 115 (MNI coordinates [-23.4,61,-6.8]), were discovered as important regions that were shared between all brain states when predicting fluid intelligence. In addition to these 3 regions,  FP region 240 and region 121 (MNI coordinates [-23.4,61,-6.8]), were shared between all brain states when predicting crystallized intelligence. Both regions 115 and 121 were not assigned to any functional module, but are located in the superior frontal gyrus of the anterior prefrontal cortex in the left hemisphere. Regions 257 and 277 were discovered as important across all brain states when predicting total composite intelligence, which have already been captured when predicting fluid and crystallized intelligence.

\begin{table}[!htbp]
\centering
\resizebox{.9\textwidth}{!}{
\begin{tabular}{|p{.23\textwidth}|p{.23\textwidth}|p{.23\textwidth}|p{.23\textwidth}|}
\hline
\multicolumn{4}{|l|}{\textit{Fluid intelligence}} \\
\hline
3 tasks & \multicolumn{3}{l|}{52 114 115 257 277} \\
MID \& SST \& rest & \multicolumn{3}{l|}{78  79  94 115 220 240 257 277} \\
MID \& nback \& rest & \multicolumn{3}{l|}{98 115 141 257 263 277 309} \\
SST \& nback \& rest & \multicolumn{3}{l|}{115 211 257 277} \\
3 tasks \& rest & \multicolumn{3}{l|}{115 257 277} \\
\hline
\multicolumn{4}{|l|}{\textit{Crystallized intelligence}} \\
\hline
3 tasks & \multicolumn{3}{l|}{78 114 115 121 240 241 257 277 320} \\
MID \& SST \& rest & \multicolumn{3}{l|}{94 115 121 150 240 257 277} \\
MID \& nback \& rest & \multicolumn{3}{l|}{115 121 140 240 257 258 277} \\
SST \& nback \& rest & \multicolumn{3}{l|}{115 121 240 252 257 277} \\
3 tasks \& rest & \multicolumn{3}{l|}{115 121 240 257 277} \\
\hline
\multicolumn{4}{|l|}{\textit{Total composite intelligence}} \\
\hline
3 tasks & \multicolumn{3}{l|}{52 167 257 277} \\
MID \& SST \& rest & \multicolumn{3}{l|}{150 257 277} \\
MID \& nback \& rest & \multicolumn{3}{l|}{98 257 258 263 277 309} \\
SST \& nback \& rest & \multicolumn{3}{l|}{211 252 257 277} \\
3 tasks \& rest & \multicolumn{3}{l|}{257 277} \\
\hline
\multicolumn{4}{|l|}{{\bf Region} information}  \\
\hline
\textit{Hemisphere} & \textit{Functional module} & \textit{Region index} & \textit{MNI coordinates} \\
\hline
\multirow{11}{*}{Left hemisphere} & Fronto Parietal & {\bf 78} & [-40.3,50.4,-4.8] \\
\cline{2-4}
 & Dorsal Attention & {\bf 52} & [-42.9,-45.0,43.0] \\
\cline{2-4}
 & \multirow{3}{*}{Default mode} &  {\bf 94} & [-39.3,-73.9,38.3] \\
 & & {\bf 114} & [-27.5,53.6,0.0] \\
 & & {\bf 150} & [-6.5,54.7,18.1] \\
\cline{2-4}
 & \multirow{3}{*}{Visual} & {\bf 98} & [-34.2,-86.6,-0.5] \\
 & & {\bf 140} & [-25.2,-97.2,-7.9] \\
 & & {\bf 141} & [-22.6,-81.7,-11.7] \\
\cline{2-4}
 & Ventral Attention & {\bf 79} & [-47.2,39.0,-9.1] \\
\cline{2-4}
 & \multirow{2}{*}{None} & {\bf 115} \\
 & & {\bf 121} & [-23.8,52.2,-12.8] \\
\hline
\multirow{12}{*}{Right hemisphere} & \multirow{4}{*}{Fronto Parietal} & {\bf 167} & [47.9,-42.5,41.5] \\
 & & {\bf 240} & [42.8,48.3,-5.1] \\
 & & {\bf 277} & [28.4,57.0,-5.1] \\
 & & {\bf 320} & [30.9,52.2,9.9] \\
\cline{2-4}
 & \multirow{2}{*}{Dorsal Attention} & {\bf 211} & [38.8,-42.6,40.4] \\
 & & {\bf 252} & [23.0,-66.4,51.8] \\
\cline{2-4}
 & \multirow{2}{*}{Default mode} & {\bf 220} & [48.9,-53.0,28.6] \\
 & & {\bf 257} & [7.4,-69.3,49.9] \\
\cline{2-4}
 & \multirow{3}{*}{Visual} & {\bf 258} & [35.4,-77.1,21.1] \\
 & & {\bf 263} & [31.7,-85.7,2.4] \\
 & & {\bf 309} & [20.4,-87.3,-6.6] \\
\cline{2-4}
 & Ventral Attention & {\bf 241} & [48.1,38.3,-9.2] \\
\hline
\end{tabular}}
\caption{Common top 10\% brain regions shared across 3 or more brain states}
\label{tab:brain_common}
\end{table}

\begin{figure}[!htbp]
\centering
\includegraphics[width=.75\textwidth]{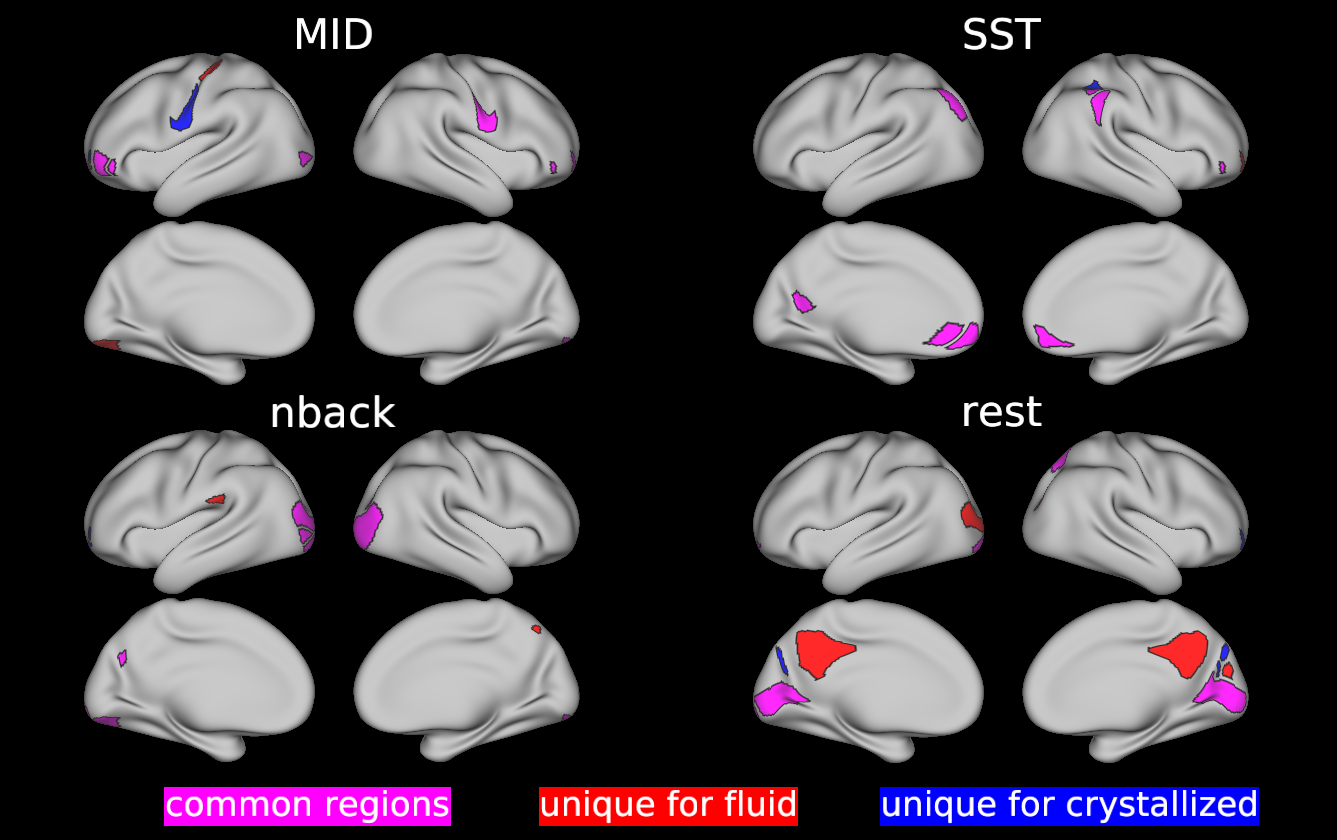}
\caption{Brain maps for top 10 shared and differential brain regions in predicting fluid and crystallized intelligence}
\label{fig:map}
\end{figure}

{\noindent \bf \underline{Shared and Differential Brain Regions across Intelligence Metrics:}}  Our previous heatmaps in Fig. \ref{fig:imp_heat} illustrated strong importance score correlations when predicting different intelligence metrics, based on a given brain state. This would suggest the presence of common brain regions that are activated when predicting different types of intelligence metrics. Table \ref{tab:features} lists the top 10 regions shared regions that show up to be important when predicting both fluid and crystallized intelligence metrics, as well as differential regions that are important for predicting fluid intelligence but not cyrstallized intelligence and vice-versa. Each of the three tasks have more shared regions with respect to fluid and crystallized intelligence, and fewer differential regions. In contrast, important brain regions at resting state are seen to be more differential when predicting fluid and crystallized intelligence. These regions are also visually illustrated in Fig. \ref{fig:map}.

The important brain regions shared across the fluid and crystallized intelligence metrics appear to be more widely distributed for the MID task (scattered across the fronto-parietal, visual, ventral attention and sensory motor networks), but they were almost exclusively concentrated within the visual network for prediction based on the nback task, and the shared regions were largely concentrated within the default mode network for prediction based on the SST task. The preponderance of shared important brain regions in the visual network when predicting intelligence metrics based on the nback task is not surprising given that this task involves visual patterns in the experiment, which evoke strong activations in the visual network that appear to drive changes in both fluid and crystallized intelligence. We note that the nback task involves working memory, and the role of visual cortex in decoding working memory has been investigated \citep{hallenbeck2021working}. For resting state, the shared important regions for predicting fluid and crystallized intelligence primarily belong to the visual network, while the distinct brain regions that show differentiation when predicting fluid and crystallized intelligence are concentrated in the visual and default mode network. These brain regions have been shown in previous literature to be connected to individual's intelligence performance \citep{gray2003neural,van2009efficiency,yuan2012salience,santarnecchi2015intelligence,song2009default}.

\begin{table}[!ht]
\centering
\resizebox{.8\textwidth}{!}{
\begin{tabular}{|p{.11\textwidth}|p{.27\textwidth}|p{.27\textwidth}|p{.27\textwidth}|}
\hline
\textit{Experiment} & \textit{Common regions} & \textit{Unique for fluid} & \textit{Unique for crystallized} \\
\hline
\multirow{8}{*}{MID} & 78: L.FP [-40.3 50.4 -4.8] & 38: L.SM [-35.8 -29.7 54.5] & 53: L.SM [-51.5 -11.9 29.7] \\
 & 79: L.VA [-47.2 39 -9.1] & 141: L.Vis [-22.6 -81.7 -11.7] & 114: L.DM [-27.5 53.6 0] \\
 & 98: L.Vis [-34.2 -86.6 -0.5] & & \\
 & 115: L.none [-23.4 61 -6.8] & & \\
 & 212: R.SM [53.9 -8.3 26.1] & & \\
 & 241: R.VA [48.1 38.3 -9.2] & & \\
 & 277: R.FP [28.4 57 -5.1] & & \\
 & 309: R.Vis [20.4 -87.3 -6.6] & & \\
\hline
\multirow{9}{*}{SST} & 13: L.RT [-14.4 -57.8 18.4] & 277: R.FP [28.4 57 -5.1] & 211: R.DA [38.8 -42.6 40.4] \\
 & 94: L.DM [-39.3 -73.9 38.3] & & \\
 & 116: L.DM [-5.9 54.8 -11.3] & & \\
 & 117: L.DM [-6.8 38.2 -9.4] & & \\
 & 151: L.DM [-15.7 64.7 13.7] & & \\
 & 167: R.FP [47.9 -42.5 41.5] & & \\
 & 219: R.CO [57.5 -40.3 34.7] & & \\
 & 241: R.VA [48.1 38.3 -9.2] & & \\
 & 279: R.DM [7.2 48.4 -10.1] & & \\
\hline
\multirow{8}{*}{nback} & 89: L.CP [-12.7 -64.9 31.8] & 103: L.CO [-55.1 -32.3 23] & 114: L.DM [-27.5 53.6 0] \\
 & 97: L.Vis [-31.3 -84.2 9] & 257: R.DM [7.4 -69.3 49.9] & 115: L.none [-23.4 61 -6.8] \\
 & 98: L.Vis [-34.2 -86.6 -0.5] & & \\
 & 140: L.Vis [-25.2 -97.2 -7.9] & & \\
 & 141: L.Vis [-22.6 -81.7 -11.7] & & \\
 & 258: R.Vis [35.4 -77.1 21.1] & & \\
 & 263: R.Vis [31.7 -85.7 2.4] & & \\
 & 309: R.Vis [20.4 -87.3 -6.6] & & \\
\hline
\multirow{6}{*}{rest} & 15: L.Vis [-11.3 -83.2 3.9] & 1: L.DM [-11.2 -52.4 36.5] & 90: L.Vis [-13.7 -77.4 26.6] \\
 & 115: L.none [-23.4 61 -6.8] & 97: L.Vis [-31.3 -84.2 9] & 255: R.Vis [17.6 -78.3 34] \\
 & 140: L.Vis [-25.2 -97.2 -7.9] & 162: R.DM [12.3 -51.6 34.5] & 277: R.FP [28.4 57 -5.1] \\
 & 175: R.Vis [15.5 -74.1 9.4] & 310: R.Vis [5.1 -80.2 23.1] & 311: R.Vis [14.6 -70.3 23.3] \\
 & 252: R.DA [23 -66.4 51.8] & & \\
 & 258: R.Vis [35.4 -77.1 21.1] & & \\
\hline
\end{tabular}}
\caption{Location information for the top 10 brain regions in predicting fluid and crystallized intelligences. Region information in `a.b [c]' format. a indicates hemisphere: L - left hemisphere, R - right hemisphere. b indicates functional module assignment. c includes the x, y, z coordinates of the region centroid in the MNI space.}
\label{tab:features}
\end{table}

\section{Discussion}
In this article, we have performed a comprehensive comparison of different types of fMRI features (static FC, dynamic FC and region level fMRI time series data) for cognitive prediction across different intelligence metrics and fMRI experiments, using the large-scale ABCD study that involved nearly 7000 individuals. We used a deep neural network involving a novel bi-LSTM approach that naturally incorporated time-varying features in the fMRI data. Our analysis illustrated distinct advantages in prediction based on time-varying fMRI features compared to prediction based on the routinely used static FC features across brain states and intelligence metrics. Our prediction results suggest the strong potential of using region level fMRI time series data for predicting intelligence metrics, particularly fluid intelligence and total composite intelligence. This is appealing considering the fast computational speeds of the bi-LSTM prediction pipeline based on region level brain features and the straightforward interpretability of these features. Additionally, dynamic FC based prediction also resulted in gains over prediction using static FC features, for crystallized intelligence. Unlike region level fMRI features, dynamic FC features are high-dimensional that may result in high computation times when combining data from multiple fMRI experiments under the bi-LSTM model. Overall, prediction using both region level fMRI data and dynamic FC features that incorporate the temporal dynamic in fMRI experiments considerably outperformed prediction based on static FC, which is currently considered the state of the art for intelligence prediction.

Another innovative and practical aspect of our analysis is the feature selection mechanism that is implemented via a $L_0$ regularization on the time-varying input features in the bi-LSTM framework. A cross-validation analysis yielded high test-retest reliability of the feature importance scores generated under this approach, which showed strong reproducibility and validated the robustness of our method. We note that while more interpretable linear regression models can also be used for feature selection, we do not use these models for analyzing important brain regions due to their poor predictive ability. Using the proposed feature selection mechanism, we provided a thorough analysis of shared and differential {\it brain regions} that are discovered as important when predicting different intelligence metrics across varying brain states. Our analysis provide complimentary results to existing literature that has shown shared and unique {\it brain network features} between task and rest fMRI that drive cognition and intelligence \citep{elliott2019general,chen2022shared}.

Both the default mode network and fronto-parietal networks are shown to contain a number of brain regions that were common across the three tasks when predicting intelligence. Of these regions, region number 257 in the default mode network and region number 277 in the fronto-parietal network were discovered as important across task and rest fMRI experiments, when predicting fluid, crystallized and total composite intelligence. Region 257 lies in the right precuneus Brodmann area 7, which has been shown to be important in visuo-motor coordination \citep{cavanna2006precuneus}. Region 277 lies in the right superior frontal gyrus of the brain, which has been associated with cognitive control of impulsive responses \citep{hu2016right}. Additionally, region number 115, which did not belong to any pre-specified functional network, was also shown to be an important region across all brain states when predicting both fluid and crystallized intelligence. This region lies in the left superior frontal gyrus of the brain that has been shown to be associated with higher cognitive functions such as working memory and spatial cognition \citep{boisgueheneuc2006functions}.

When predicting intelligence based on the nback task fMRI time series, several shared regions in the visual network were discovered that were important for predicting both fluid and crystallized intelligence. These findings are consistent given that the nback task involves visual stimuli and is expected to recruit regions in the brain involved in visual processing. The specific brain regions in the visual network that were recruited during the nback task and are important for predicting both fluid and crystallized intelligence include regions 97 and 98 in the left middle occipital gyrus, region 140 in the left inferior occipital gyrus, and region 141 in the left lingual gyrus, which are all parts of the secondary visual cortex in the left hemisphere. Additional important regions include regions 258 and 263 in the right middle occipital gyrus, and region 309 in the right lingual gyrus, which all belong to the secondary visual cortex in the right hemisphere. These findings are consistent with existing literature that have discovered the importance of occipital regions \citep{miro2020locating} and the visual cortex \citep{hallenbeck2021working} in the context of the nback task.

On the other hand, shared regions that were common for predicting fluid and crystallized intelligence based on the SST task involved several brain areas from the default mode network. In the SST task, the participant responds to an arrow stimulus, by selecting one of two options, depending on the direction in which the arrow points. If an audio tone is present, the subject must withhold making that response (inhibition). The test consists of an initial practice phase, and a subsequent task phase where the auditory stop signal is generated according to some design unknown to participants. Given that the SST task involves a learning (practice) phase, and transitions between unknown stops and task, the recruitment of the default mode network in the SST task that is associated with cognitive ability is supported by previous literature. Existing work shows that the DMN is recruited in switching tasks, in the case of a demanding shift from a cognitive context to a different one \citep{crittenden2015recruitment}, as well as during decision-making \citep{smith2021roles} and when subjects have to automatically apply learned rules \citep{vatansever2017default}. In contrast, although it is known that the default mode network is more active during rest compared to task, none of the brain regions in the default mode network during rest are found to be strongly associated with fluid or cyrstallized intelligence. This implies that the DMN at rest is not pertinent for predicting cognitive abilities, which is not surprising give existing literature that suggests that the DMN during rest may be associated with self-generated thoughts and mind-wandering \citep{andrews2014default,christoff2016mind}.

\section{Conclusion}
In this work, our comprehensive comparison of predictive abilities across different types of fMRI features and involving several task and rest fMRI experiments, adds to the limited understanding in the current literature. Our investigations highlight the importance in accounting for the temporal variations in fMRI features when predicting the fluid, crystallized and total composite intelligence, and further showcase the advantages of using region level fMRI time series compared to network-based analysis in terms of computational efficiency and interpretability. The proposed bi-LSTM model with feature selection mechanism shows great potential for scientific discoveries in neuroimaging applications with fMRI experiments. Our source code for the proposed method is available online at \url{https://github.com/leo-yangli/abcd_time_series}.

\singlespacing
\bibliographystyle{apalike}
\bibliography{brain}
\end{document}